\documentclass{article}
\usepackage{spconf}
\usepackage{times}  
\usepackage{helvet} 
\usepackage{courier}  
\usepackage[hyphens]{url}  
\usepackage{graphicx} 
\usepackage{siunitx}
\usepackage{amsfonts}
\usepackage{caption}
\DeclareCaptionStyle{ruled}{labelfont=normalfont,labelsep=colon,strut=off} 
\usepackage{algorithm}
\usepackage{algorithmic}
\usepackage{amssymb}
\usepackage{booktabs, tabularx, colortbl}
\usepackage{multirow}
\usepackage{makecell}
\usepackage{amsmath}
\usepackage{subfig}
\usepackage{xcolor}
\newcommand{\tabitem}{~~\llap{\textbullet}~~}
\usepackage{pifont}
\usepackage{hyperref}
\usepackage{newfloat}
\usepackage{listings}
\floatstyle{ruled}
\newfloat{listing}{tb}{lst}{}
\floatname{listing}{Listing}
\newcommand{\mcrot}[4]{\multicolumn{#1}{#2}{\rlap{\rotatebox{#3}{#4}~}}} 

\title{Blind Estimation of Audio Processing Graph}
\name{Sungho Lee$^{1}$, Jaehyun Park$^{1}$, Seungryeol Paik$^{1}$, and Kyogu Lee$^{1,2,3}$
\thanks{Project page: \href{https://sh-lee97.github.io/apg}{\texttt{https://sh-lee97.github.io/apg}}. \hfil\break 
\textbf{Acknowledgement.} This work was supported by Culture, Sports, and Tourism R\&D Program through the Korea Creative Content Agency grant funded by the Ministry of Culture, Sports and Tourism in 2022 (No.R2022020066).}}
\address{
$^{1}$Department of Intelligence and Information, Seoul National University\\ 
$^{2}$Interdisciplinary Program in Artificial Intelligence, Seoul National University\\ $^{3}$Artificial Intelligence Institute, Seoul National University\\
\texttt{\{sh-lee,lotussoh,paik402,kglee\}@snu.ac.kr}}

\begin{document}
\ninept

\maketitle
\begin{abstract}
Musicians and audio engineers sculpt and transform their sounds by connecting multiple processors, forming an \emph{audio processing graph}. 
However, most deep-learning methods overlook this real-world practice and assume fixed graph settings.
To bridge this gap, we develop a system that reconstructs the entire graph from a given reference audio. 
We first generate a realistic graph-reference pair dataset and train a simple blind estimation system composed of a convolutional reference encoder and a transformer-based graph decoder.
We apply our model to singing voice effects and drum mixing estimation tasks. Evaluation results show that our method can reconstruct complex signal routings, including multi-band processing and sidechaining.
\end{abstract}
\vspace{-2.5mm}
\section{Introduction}
\begin{figure}
    \begin{center}
        \includegraphics[width=\columnwidth]{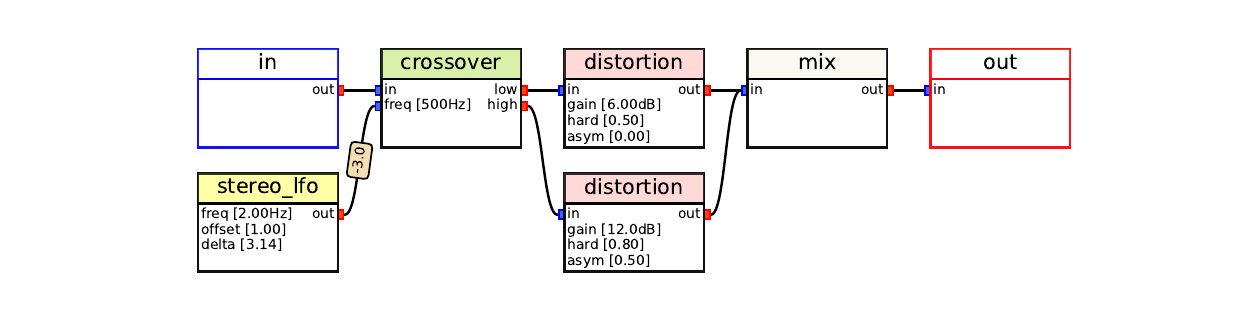}
        \vspace{-4mm}
        \caption{An example audio processing graph $G$.} 
        \label{fig:graph}
    \end{center}
    \vspace{-9mm}
\end{figure}

\vspace{-1.5mm}
A dry audio signal is rarely delivered to the end listeners ``as is"--- it goes through various processing steps to achieve desired auditory effects.  
Recent deep learning approaches have improved, replaced, and automated some parts of such processing pipelines with neural networks. Some focused on estimating the parameters of known conventional DSP systems \cite{ramirez2020blackbox, Mitcheltree_2021, engel2020ddsp, colonel2021reverse}, while others mimicked, augmented, or integrated the existing processors with neural networks \cite{ramirez2019general, neuralbiquads, steinmetz2020diffmixconsole}.  
These prior works consider one or a few types of processors with fixed signal routings. 
However, domain practitioners, e.g., musicians and audio engineers, typically connect multiple small audio processors in a flexible way to achieve the desired processing. 
If we consider these processors as nodes and the connections as edges, we can represent this procedure as an \emph{audio processing graph} $G$.
See Figure \ref{fig:graph} for an example. First, we feed an input dry signal $x$, e.g., singing voice, into the \texttt{[in]} node. Then, the \texttt{[crossover]} splits the signal into two, one with low-frequency components and the other with high-frequency components (written as \texttt{low} and \texttt{high}, respectively). The two signals are independently processed with their respective \texttt{[distortion]} modules, summed together with \texttt{[mix]}, and passed to the \texttt{[out]} node, resulting in an output signal $y=G(x)$. This graph $G$ performs so-called ``multiband distortion," and we can extend its functionality with additional processors and connections. For example, we can add \texttt{[stereo\_lfo]} to modulate the crossover frequency, making the distortion effect time-varying. This flexibility makes expressive processing possible while retaining full interpretability and controllability.

In this paper, we conduct a preliminary study on integrating the audio processing graph structure with neural networks. Specifically, we tackle a \emph{blind estimation} problem; from the reference audio $y$, we aim to reconstruct the original graph $G$. 
Our motivation for choosing this task is threefold. First, automated reverse engineering itself is a well-established task and has practical values \cite{colonel2021reverse}. Second, it provides concrete objective metrics, allowing us to evaluate baseline systems reliably. Finally, it could serve as an appropriate first step towards applying and extending the graph-based methods to other applications, which include automatic processing \cite{martinez2022automatic, koo2022remaster, stasis2017audio} and style transfer \cite{steinmetz2022styletransfer}.

The blind graph estimation task poses several challenges to overcome. First, there is no publicly available graph-audio paired dataset. Therefore, we first build a \emph{synthetic dataset} on our own. Second, audio processing graphs are heterogeneous; each node and edge have a type and parameters as attributes, making them nontrivial to decode. To mitigate this, we propose a two-stage decoding method consisting of (i) \emph{prototype graph decoding}, which estimates graph connectivity and node/edge types, and (ii) \emph{parameter estimation}, which decides the remaining parameters. We achieve this with a convolutional reference encoder \cite{hershey2017cnn} and a transformer-based graph decoder \cite{vaswani2017attention, kim2022pure}. 

Finally, we apply our method to singing voice effects and drum mixing estimation tasks. The evaluation results show that, while our model fails to perfectly reconstruct the original graph in most cases, it preserves essential local/global graph structures, and rendered audio from the estimated graph can be perceptually similar to the reference. We also show that the proposed graph decoding strategy is preferable to other possible alternatives.

\vspace{-1.5mm}
\section{Audio Processing Graph}
\vspace{-1.5mm}

\begin{table}[t]
\setlength\tabcolsep{3.8pt}
\renewcommand{\arraystretch}{.78}
\begin{center}
\fontsize{8.1}{8.1}\selectfont
\begin{tabular}{l}
\toprule
{ \textit{Processor(s):} {\color{blue}[inlets, optional$^*$]} $\to$ {\color{red}[outlets];} {\color{teal}[parameters].}}\\
\midrule
\arrayrulecolor{darkgray} 
\textit{Low-order linear filters} \cite{vafilter} \vspace{.35mm}\\
{\it \tabitem Second-order low/band/highpass, bandreject, and fourth-order}\\
\quad{\it low/band/highpass:}
{\color{blue}[in, frequency$^*$]} $\to$ {\color{red}[out];} {\color{teal}[frequency, q].}
\vspace{.35mm}\\ 
{\it \tabitem Parametric equalizer filters - low/highshelf and bell (peaking filter):}\\
\quad{\color{blue}[in, frequency$^*$, gain$^*$]} $\to$ {\color{red}[out];} {\color{teal}[frequency, q, gain].}\vspace{.35mm}\\
{\it \tabitem Crossover:} {\color{blue}[in, frequency$^*$]} $\to$ {\color{red}[low, high];} {\color{teal}[frequency].\vspace{.35mm}}\\
{\it \tabitem Phaser:} {\color{blue}[in, mod]} $\to$ {\color{red}[out];} {\color{teal} [frequency, feedback, mix].}\\
\midrule
\textit{High-order linear filters} \cite{smith2010physical}\vspace{.35mm}\\
{\it \tabitem Chorus/flanger/vibrato:} {\color{blue}[in, mod]} $\to$ {\color{red}[out];} {\color{teal}[delay, feedback, mix].}\vspace{.35mm}\\ 
{\it \tabitem Mono and pingpong delay:} {\color{blue}[in]} $\to$ {\color{red}[out];}\\
\quad
{\color{teal}[delay, feedback, mix, frequency, q, stereo\_offset].}\vspace{.35mm}\\ 
{\it \tabitem Reverb (mono and stereo):} {\color{blue}[in]} $\to$ {\color{red}[out];} {\color{teal}[size, damping, width, mix].}\\ 
\midrule
\textit{Nonlinear filters} \vspace{.35mm}\\
{\it \tabitem Distortion} \cite{eichas2020system}: 
{\color{blue}[in]} $\to$ {\color{red}[out];} {\color{teal}[gain, hardness, asymmetry].}\!\vspace{.35mm}\\
{\it \tabitem Bitcrush:} {\color{blue}[in]} $\to$ {\color{red}[out];} {\color{teal}[bit].}\vspace{.35mm}\\
{\it \tabitem Dynamic range controllers - compressor/noisegate/expander} \cite{giannoulis2012digital}:\\
\quad{\color{blue}[in, sidechain$^*$]} $\to$ {\color{red}[out];} {\color{teal}[threshold, ratio, attack, release, knee].}\vspace{.35mm}\\
{\it \tabitem Pitchshift}: {\color{blue}[in]} $\to$ {\color{red}[out];} {\color{teal}[semitone].}\\
\midrule
\textit{Utility processors}\vspace{.35mm}\\
{\it \tabitem Mix:} {\color{blue}[in]} $\to$ {\color{red}[out];} {\color{teal}[].}\vspace{.35mm}\\
{\it \tabitem Panning:} {\color{blue}[in, pan$^*$]} $\to$ {\color{red}[out];} {\color{teal}[pan].}\vspace{.35mm}\\
{\it \tabitem Imager:} {\color{blue}[in]} $\to$ {\color{red}[out];} {\color{teal}[width].}\vspace{.35mm}\\
{\it \tabitem Mid/side splitter:} {\color{blue}[in]} $\to$ {\color{red}[mid, side];} {\color{teal}[].}\vspace{.35mm}\\
{\it \tabitem Mid/side merger:} {\color{blue}[mid, side]} $\to$ {\color{red}[out];} {\color{teal}[].}\\
\midrule
\textit{Control signal generators}\vspace{.35mm}\\
{\it \tabitem Low-frequencyuency oscillator (mono and stereo):}\\
\quad{\color{blue}[]} $\to$ {\color{red}[lfo];} {\color{teal}[frequency, phase, stereo\_offset].}\vspace{.35mm}\\ 
{\it \tabitem Envelope follower:} 
{\color{blue}[in]} $\to$ {\color{red}[env];} {\color{teal}[attack, release, gain].}\\
\arrayrulecolor{black} 
\bottomrule
\vspace{-3.3mm}
\end{tabular}
\caption{List of the processors and their configurations.}
\vspace{-1mm}
\label{table:processors}
\end{center}
\vspace{-8.6mm}
\end{table}

Our audio processing graph $G$ is a heterogeneous directed acyclic graph (DAG) with the following specifications. 

\noindent \textbf{Processors/Nodes.} Each processor $v_i$ can take audio/control signals as input and output audio/control signals. We normalize each processor's output audio signals so that the total energy remains unchanged.
Each node has a categorical type $t_i$ and continuous-valued parameters $p_i$ as attributes. 
While we can include any processor (even neural networks) in our framework, we use $33$ conventional ones listed in Table \ref{table:processors}. Unless stated otherwise, we use the default implementations \cite{zolzer2011dafx}.
This makes our graph readily interpretable and controllable.

\noindent \textbf{Connections/Edges.} Each connection $e_{ij}$ requires outlet $m_{ij}$ and inlet $n_{ij}$ (or input/output channel) attributes to eliminate any ambiguity. We define the edge's type $t_{ij}$ as an outlet-inlet pair $(m_{ij}, n_{ij})$. When multiple edges are connected to the inlet, we sum the incoming signals. Each edge has a gain parameter $p_{ij}$. 

\vspace{-3mm}
\subsection{Synthetic Graph for Training}
\begin{figure}
    \centering
    \vspace{-1.3mm}
    \subfloat[Statistics of the singing and drum graphs compared to the PCQM4Mv2. 
    \label{fig:graph-stats}]{
        \includegraphics[width=.96\columnwidth]{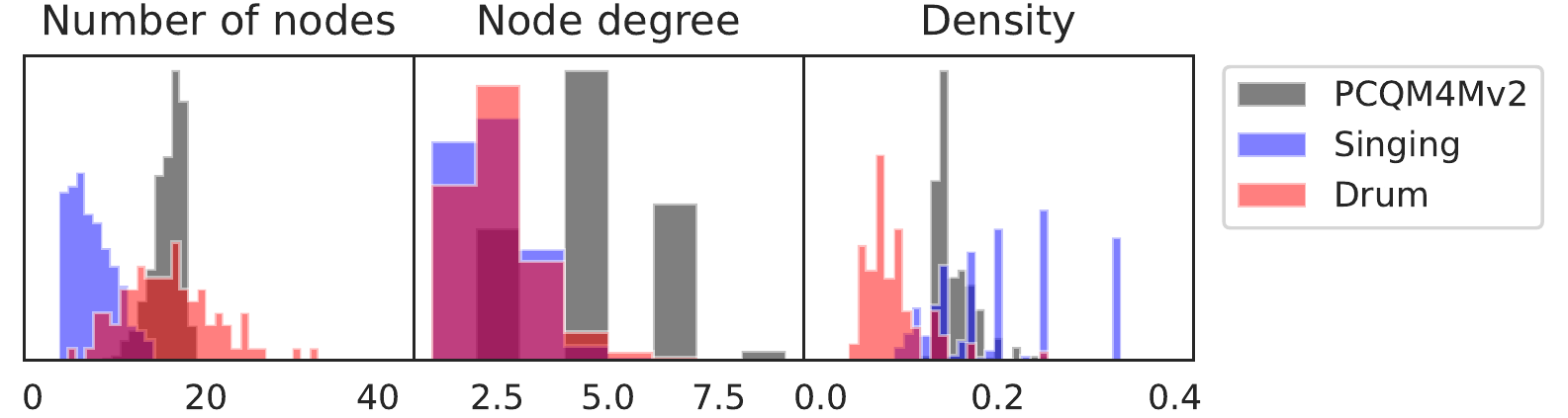}\vspace{-6.7mm}}\vspace{-5.5mm} \\
    \subfloat[Reordering of the LTI (sub)graphs.
    \label{fig:lti-reordering}]{
        \includegraphics[width=.75\columnwidth]{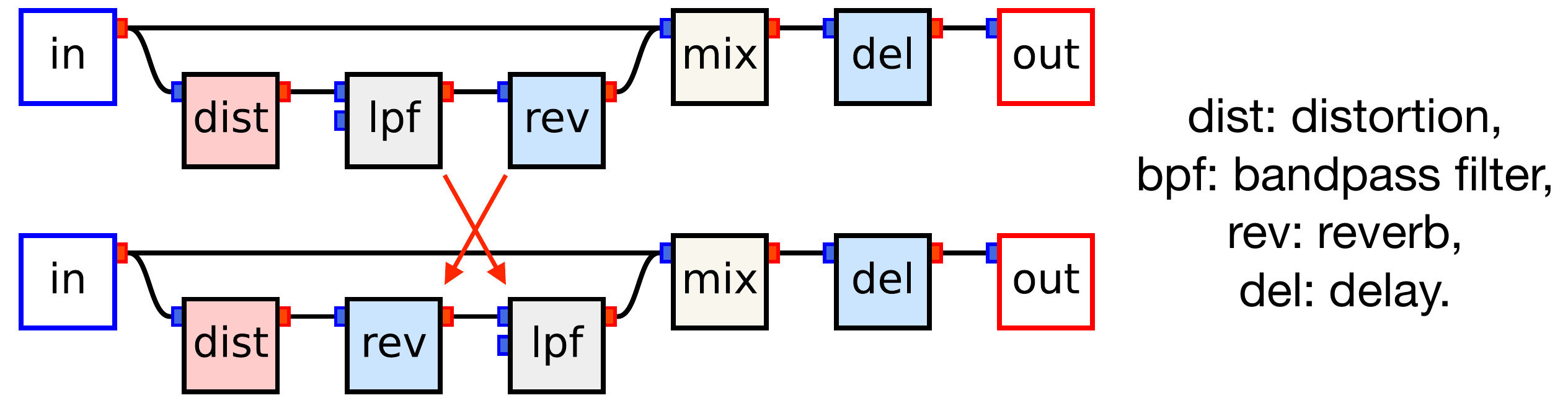}}\vspace{-2mm} \\
\caption{Properties of audio processing graph:}
\vspace{-5mm}
  \label{fig:properties} 
\end{figure}

\vspace{-1.5mm}
\noindent \textbf{Prototype Graph.}
The real-world audio processing graphs have frequently occurring \emph{motifs}: combinations of processors (subgraphs) that achieve desired effects. For example, we can control \texttt{[reverb]} using \texttt{[noisegate]}, resulting in the well-known ``gated reverb." 
Inspired by this, we sample and combine various motifs to obtain a synthetic graph. We have $10$ different motifs, from a simple parametric equalizer to more complex parallel $\texttt{[pitchshift]}$ banks. The sampled motifs are serially stacked to generate a full graph except for the following cases: (i) \texttt{[crossover]} can be used for multi-band processing and 
(ii) each motif can receive auxiliary signals from the others for various modulations, e.g., sidechaining. For the drum mixing graphs, we generate a subgraph for each individual source track. Then, we combine the subgraphs with another ``mixing bus" graph. See Figure \ref{fig:lti-reordering}, \ref{fig:drum-est}, and \ref{fig:singing-est} for examples of the synthesized graphs. In this stage, we only determine categorical type attributes $t_i$ and $t_{ij}$ of the graph, and we call this a \emph{prototype graph} $G_0$.

\noindent \textbf{Adaptive Parameter Randomization.} 
Next, we decide the remaining parameters of each prototype graph.
Specifically, we randomize them adaptively to the incoming signals, preventing the graph from having ``ghost" nodes that do not contribute to the final output signal. 
For example, when a \texttt{[highshelf]} receives a \texttt{[crash]} signal, its center frequency should be constrained to the high-frequency region to change the frequency response of the input audio.
To achieve this, we compute the cumulative energy distribution of the input signal across the frequency, then sample the center frequency from where the cumulative distribution lies between $0.2$ and $0.8$. We follow similar procedures to the other low-order linear filters. For the dynamic range controllers, we use an input energy envelope to determine their thresholds. This way, we generated $300\si{k}$ and $450\si{k}$ graph-reference audio pairs for the singing and drum, respectively. The reference audio is stereo, $3.63\si{s}$ long, and has $44.1\si{kHz}$ sampling rate. 

\noindent \textbf{Graphs Statistics.}
Figure \ref{fig:graph-stats} reports the graph statistics of our synthetic graphs.
Especially, we compare our datasets with PCQM4Mv2 chemical dataset \cite{nakata2017pubchemqc},
While the singing/drum graphs have a comparable size to the PCQM4Mv2's, they tend to be more sparse (lower node degree and density). This implies that simple graph neural networks which update each node via aggregating itself and its immediate neighborhoods might struggle to make distant nodes communicate and capture the global structures of the audio processing graphs.

\noindent \textbf{LTI Reordering.}
Different audio processing graphs can produce the same output, making the blind graph estimation a one-to-many problem. One reason is that serially connected single-input single-output linear time-invariant (LTI) systems, either single nodes or subgraphs, can be swapped without changing the entire system response. To resolve this ambiguity, we rearrange them according to their sizes and processor types (see Figure \ref{fig:lti-reordering}). 

\vspace{-2mm}
\section{Blind Estimation System}
\begin{figure}[!h]
    \vspace{-1.9mm}
    \begin{center}
        \includegraphics[width=.92\columnwidth]{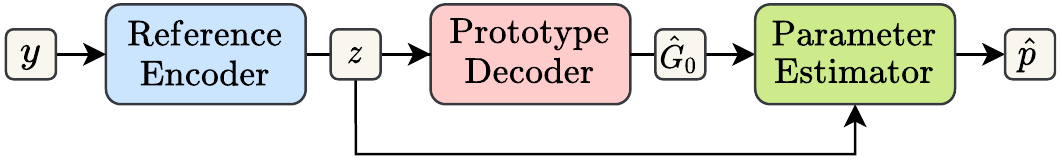}
        \vspace{-2mm}
        \caption{The proposed blind estimation framework.} 
        \label{fig:framework}
    \end{center}
    \vspace{-6mm}
\end{figure}

\vspace{-2mm}
Our blind estimation system first encodes the reference $y$ into latent vectors $z$, which should contain the necessary information to estimate the graph. Then, from the latent $z$,  we reconstruct the graph in two stages, which resemble the synthetic data generation procedure; we first decode the prototype $\smash{\hat{G}_0}$ autoregressively, then estimate the remaining parameters $\hat{p}$ (see Figure \ref{fig:framework}).

\noindent \textbf{Reference Encoder.} 
We first apply $7$ two-dimensional convolutional layers to the reference Mel spectrogram.
Then, we flatten and project the channel and frequency axes of the output feature map into $512$-dimensional vectors. 
Finally, we perform an attentive pooling across the time axis with learnable queries to obtain latent vectors $z$. 

\begin{figure}
    \begin{center}
        \includegraphics[width=.97\columnwidth]{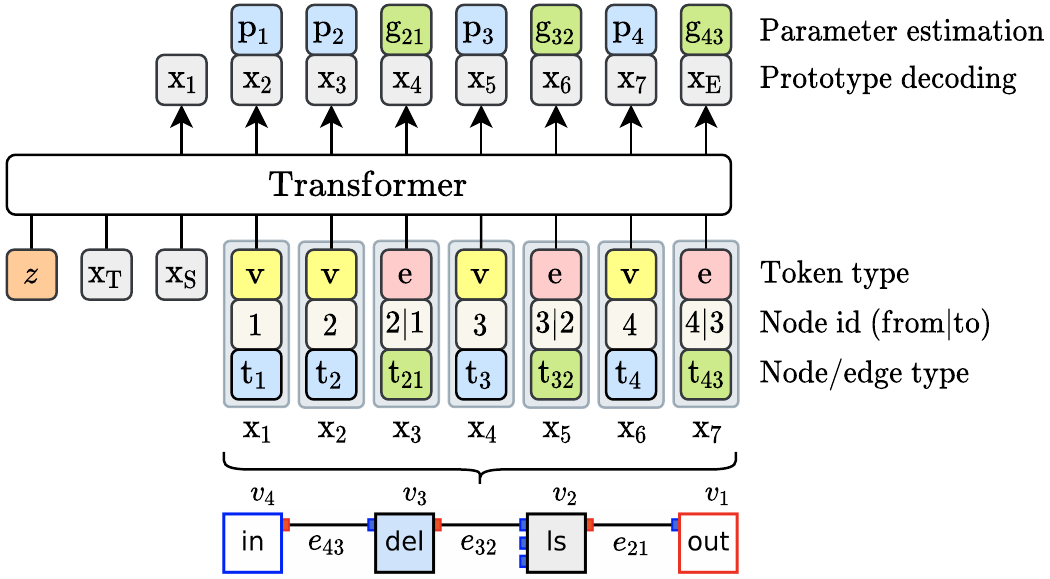}\vspace{-2mm}
        \caption{Blind estimation with Tokenized Graph Transformer.} 
        \label{fig:tokengt}
    \end{center}
    \vspace{-9.5mm}
\end{figure}

\noindent \textbf{Prototype Decoder.}
Most autoregressive models \cite{faez2021deep, li2018learning, liao2019efficient} generate graphs \emph{node-by-node}; for each step, they observe a partially decoded subgraph, then estimate the next node and its edges.
Instead, we present an alternative method called \emph{token-by-token} generation with Tokenized Graph Transformer (TokenGT, see Figure \ref{fig:tokengt}) \cite{kim2022pure}. TokenGT uses a vanilla transformer \cite{vaswani2017attention} and treats both nodes and edges as input tokens. 
It introduces token type embeddings that distinguish the nodes and edges and node id embeddings to describe the connectivity (we additionally have node/edge type embeddings).
This model architecture allows us to (i) alleviate the potential information bottleneck problem due to the sparse audio processing graph and (ii) frame the prototype decoding as a canonical autoregressive sequence generation task.
To decode the graph, we first add a start-of-graph token $\mathrm{x_S}$ to the empty sequence. Then, we estimate the following graph tokens $\mathrm{x}_1, \cdots, \mathrm{x}_N$ one by one with prediction heads for the token type, node id, node type, and edge type. 
The decoding starts with the \texttt{[out]} node, then follows the breadth-first search (BFS) order. For the drum graphs, we decode the mixing subgraph first; then, we decode each individual source track subgraph one by one in BFS order. We finish the decoding when the token type estimator outputs an end-of-graph token $\mathrm{x_E}$. To condition the reference latent $z$, we concatenate it with the other input tokens. 

\noindent \textbf{Parameter Estimator.} 
We reuse the prototype decoder for the parameter estimation; we add another projection head, append a task token $\mathrm{x}_\mathrm{T}$ to differentiate the two tasks, and remove the causal attention mask. Since each parameter has a different range and scale, we translate and rescale the ground-truth value to fit into $[0, 1]$ range.

\noindent \textbf{Architecture Details.} 
The FFT size, hop length, and the number of Mel filter banks of the reference Mel spectrogram are $1536$, $384$, and $256$, respectively. The convolutional backbone is a VGGish model \cite{hershey2017cnn} with the following modifications: (i) depthwise separable convolutions \cite{chollet2017xception}, (ii) channels divided into four groups with dilations of $[1$, $2$, $4$, $8]$, (iii) and the use of layer normalization \cite{ba2016layer}.
We used a transformer decoder layer with $1$ (singing) and $6$ (drum) queries for the pooling.
For the graph decoder, we use the $6$-layer transformer encoder with the pre-layer normalization \cite{xiong2020layer} and $16$ heads. While the original paper used eigenvectors of the normalized Laplacian matrix as node id encoding, we use the sinusoidal embeddings since the eigenvectors are intractable during the decoding.

\noindent \textbf{Training.} We train the prototype decoding task in a teacher-forcing manner using the cross-entropy losses with label-smoothing of $0.1$. At the same time, we train the parameter estimation task by feeding an oracle prototype $G_0$ as input and using the $l_1$ distance as an objective. We use AdamW \cite{loshchilov2017decoupled} optimizer, a linear learning rate scheduler with \texttt{5e-4} peak learning rate, $50\si{k}$ warmup steps, $200\si{k}$ total training steps, and batch size of $32$.

\vspace{-2.5mm}
\section{Experiments}
\vspace{-3mm}
\subsection{Data}
\vspace{-1mm}
\noindent \textbf{Singing.} We used the OpenSinger dataset \cite{huang2021multi}, which has $50\si{h}$ clean recordings from $76$ speakers. We used 90\% of the audio from the $71$ speakers for the training. We use the remaining 10\% and the other $5$ speakers' recordings as two separate validation sets, \emph{seen} and \emph{unseen} speaker datasets, respectively, to compare the effect of dry source distribution shift on the model performance.

\noindent \textbf{Drum.} We collected the source signals by ourselves; we rendered the \texttt{[kick]}, \texttt{[snare]}, \texttt{[hat]}, \texttt{[tom]}, \texttt{[ride]}, and \texttt{[crash]} tracks separately with $14$ commercial sampling libraries and MIDI files from the Groove MIDI Dataset \cite{groove2019}. 
We performed equalization to the tracks so that each instrument has the same average frequency response across the kit. 
To generate each reference audio, we sampled a random segment from the dry tracks. Then, we generated a graph with the input nodes corresponding to non-zero energy tracks so that there are no dummy subgraphs. This indicates that our system should also perform a drum instrument recognition task. We used the $11$ kits and the MIDI files from the $\texttt{drummer1/session1-3}$ subset for the training. We used $\texttt{drummer1/eval\_session}$ for the two validation sets; we used the same kits for a \emph{seen} kit validation set and the remaining $3$ kits for an \emph{unseen} kit validation set. 
\begin{table*}[t]
\setlength\tabcolsep{1.94pt}
\renewcommand{\arraystretch}{.78}
\begin{center}
\fontsize{8.1}{8.1}\selectfont
\begin{tabular}{lc|rrrrrrrrc|rrrrrrrrc}
\toprule
& & \multicolumn{9}{c|}{Singing voice effect estimation} & \multicolumn{9}{c}{Drum mixing graph estimation} \\
\cmidrule{3-20}
Methods & & \mcrot{1}{l}{25}{Node error rate \!\!} &
\mcrot{1}{l}{25}{Edge error rate \!\!} &
\mcrot{1}{l}{25}{Invalid rate \!} &
\mcrot{1}{l}{25}{IOU \!} &
\mcrot{1}{l}{25}{MSS-default \!} &
\mcrot{1}{l}{25}{Parameter loss \!\!} &
\mcrot{1}{l}{25}{MSS-oracle \!} &
\mcrot{1}{l}{25}{MSS-full \!} &
MUSHRA  &
\mcrot{1}{l}{25}{Node error rate \!\!} &
\mcrot{1}{l}{25}{Edge error rate \!\!} &
\mcrot{1}{l}{25}{Invalid rate \!} &
\mcrot{1}{l}{25}{IOU \!} &
\mcrot{1}{l}{25}{MSS-default \!} &
\mcrot{1}{l}{25}{Parameter loss \!\!} &
\mcrot{1}{l}{25}{MSS-oracle \!} &
\mcrot{1}{l}{25}{MSS-full \!} &
MUSHRA  \\
\midrule

Hidden reference & 
& $-$    & $-$    & $-$    & $-$    & $-$    & $-$    & $-$    & $-$    & $92.0_{\pm 3.4}$ 
& $-$    & $-$    & $-$    & $-$    & $-$    & $-$    & $-$    & $-$    & $90.0_{\pm 3.2}$\\

\midrule

\ding{192} Autoencoding &
& $0$    & $.002$ & $.005$ & $1$    & $.002$ & $.008$ & $.004$ & $.004$ & $85.7_{\pm 5.0}$ 
& $0$    & $.001$ & $0$    & $1$    & $.001$ & $.002$ & $.018$ & $.018$ & $90.3_{\pm 4.8}$ \\

\ding{193} No latent conditioning & 
& $.502$ & $.014$ & $0$    & $.365$ & $.116$ & $.200$ & $.095$ & $.107$ & $22.1_{\pm 4.5}$ 
& $.602$ & $.024$ & $0$    & $.267$ & $.203$ & $.207$ & $.138$ & $.187$ & $32.8_{\pm 8.0}$ \\
\midrule

\ding{194} Token-by-token, two-stage & S 
& $.215$ & $.007$ & $0$    & $.839$ & $.044$ & $.120$ & $.056$ & $.058$ & $76.6_{\pm 4.6}$
& $.335$ & $.019$ & $0$    & $.659$ & $.100$ & $.145$ & $.080$ & $.085$ & $69.8_{\pm 5.4}$ \\

& U
& $.242$ & $.009$ & $0$    & $.800$ & $.054$ & $.125$ & $.057$ & $.059$ & $75.6_{\pm 4.7}$
& $.446$ & $.015$ & $0$    & $.530$ & $.159$ & $.161$ & $.088$ & $.119$ & $46.4_{\pm 5.7}$ \\

\ding{195} \ding{194} with source conditioning & U 
& $.169$ & $.011$ & $0$    & $.906$ & $.034$ & $.123$ & $.054$ & $.052$ & $77.5_{\pm 5.1}$
& $.411$ & $.019$ & $0$    & $.550$ & $.137$ & $.160$ & $.092$ & $.110$ & $59.3_{\pm 5.9}$ \\


\midrule

\ding{196} Node-by-node, two-stage & U 
& $.267$ & $.052$ & $.072$ & $.742$ & $.064$ & $.142$ & $.065$ & $.066$ & $60.9_{\pm 6.0}$
& $.461$ & $.045$ & $.132$ & $.443$ & $.134$ & $.193$ & $.106$ & $.134$ & $39.2_{\pm 5.3}$ \\

\ding{197} Token-by-token, single stage & U 
& $.249$ & $.020$ & $.002$ & $.799$ & $.053$ & $.168$ & $-$    & $.070$ & $62.4_{\pm 5.6}$
& $.442$ & $.023$ & $.023$ & $.483$ & $.159$ & $.213$ & $-$    & $.144$ & $40.6_{\pm 5.3}$ \\

\bottomrule
\vspace{-3.5mm}
\end{tabular}
\caption{Evaluation results of the proposed method and other baselines. U and S denote unseen and seen speaker/kit validation set, respectively. }
\vspace{-5mm}
\label{table:results}
\end{center}
\end{table*}

\vspace{-3mm}
\subsection{Metric}
\vspace{-1.5mm}
\noindent \textbf{Prototype Decoding.}
For each decoding step, we evaluate the node and edge error rate, counting each prediction as an error if either token type, node id, or node/edge type is incorrect.
Using the following metrics, we also compare the decoded graph with the ground truth. (i) Invalid graph rate: we consider a graph invalid if it is cyclic, not connected, or missing necessary connections. (ii) Intersection-over-union (IOU) of the node types: while this metric ignores the graph structure, it checks whether necessary processors and input nodes are decoded somewhere in the graph. For the drum graph, we calculate the IOU for each track and mixing subgraph and average the values. Finally, (iii) we render the ground-truth graph and the estimated prototype graph with default parameters and compare the outputs using multi-scale spectral loss (MSS-default) \cite{engel2020ddsp}. 

\noindent \textbf{Parameter Estimation.} Along with the parameter loss, we evaluate the MSS loss rendered on the oracle prototype with estimated parameters (MSS-oracle) and the fully-decoded graph (MSS-full). 

\noindent \textbf{Listening Test.}  We measured subjective scores with MUltiple Stimuli with Hidden Reference and Anchor (MUSHRA) test \cite{mushra, schoeffler2018webmushra}. 
We asked $8$ graduate students to score the similarity between the reference and the rendered audio with the estimated graph. A total of $48$ sets were scored ($24$ sets for each task and $12$ sets for each \emph{seen} and \emph{unseen} speaker/kits).

\begin{figure*}[t]
    \begin{center}
        \includegraphics[width=2\columnwidth]{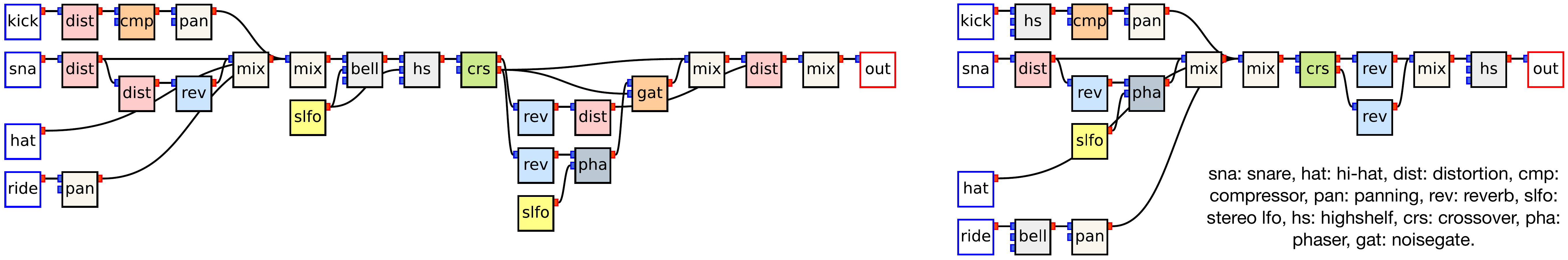}
        \vskip -2pt
        \vspace{-1mm}
        \caption{An inference result from the proposed method \ding{194} on the drum mixing estimation task (left: ground-truth, right: prediction). 
        First, our model correctly predicted the source instrument types: \texttt{[kick]}, \texttt{[snare]}, \texttt{[hat]}, and \texttt{[ride]}. The ground-truth graph panned \texttt{[kick]} and \texttt{[ride]}, which is also reconstructed by the prediction. However, the prediction failed to estimate (i) the correct multi-band processing of \texttt{[reverb]}, (ii) multiple stages of \texttt{[distortion]} processors, and (iii) modulation of linear filters in the mixing subgraph.
        } 
        \label{fig:drum-est}
    \end{center}
    \vspace{-8.5mm}
\end{figure*}

\vspace{-3mm}
\subsection{Evaluation Results}
\vspace{-1.5mm}

\noindent \textbf{Sanity Check.} 
Table 2 reports the evaluation results. Before training the blind estimation models, we first trained a \ding{192} graph autoencoder by introducing another TokenGT as a graph encoder (we also embedded the parameters for the encoder input). Its evaluation results, e.g., $0$ node error rate, confirm that the graph decoder is powerful enough and the dimension of the latent $z$ is sufficiently large to reconstruct the original graph. Furthermore, its MUSHRA score is comparable to the hidden reference's, agreeing with the objective metrics. Next, we evaluated the \ding{193} graph decoder with latent vectors set to $0$. Its node error rate ($0.502$ and $0.602$) is better than what a random guess would achieve. This is because (i) the probability distribution of the processors is nonuniform, (ii) we sorted the LTI subgraphs, and (iii) the ground-truth intermediate prototype is available to the network, which can be exploited for a better guess.

\begin{figure}[t]
    \centering
    \vspace{1.2mm}
    \subfloat[
    While our model successfully predicted that the input is processed with a  {\texttt{{[pingpong\_delay]}}} and two parallel {\texttt{[pitchshift]}} modules (each with panning modulation), it failed to estimate their correct order.
    \label{fig:singing-est-1}]{
        \includegraphics[width=.97\columnwidth]{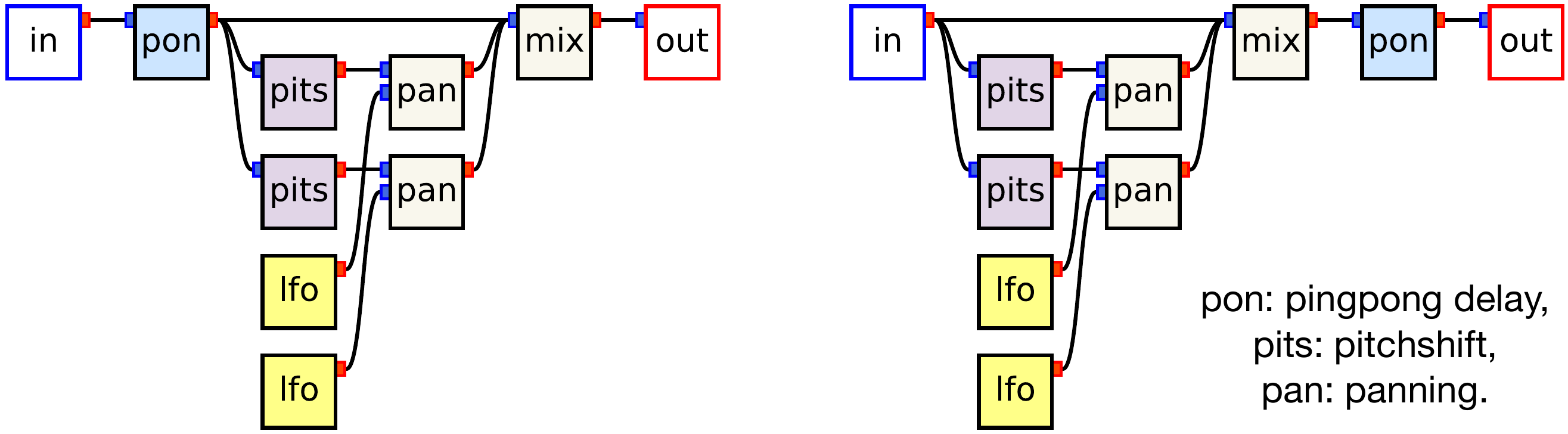}\vspace{-4.5mm}}\vspace{-.8mm} \\
    \subfloat[
    Our model failed to decode the {\texttt{[bell]}} and {\texttt{[lowshelf]}}. This is not surprising, because if {\texttt{bit}} parameter of the {\texttt{[bitcrush]}} is set to very low value, it destroys the signal and eventually makes estimation of the processors applied before the {\texttt{[bitcrush]}} challenging.
    \label{fig:singing-est-2}]{
        \includegraphics[width=.99\columnwidth]{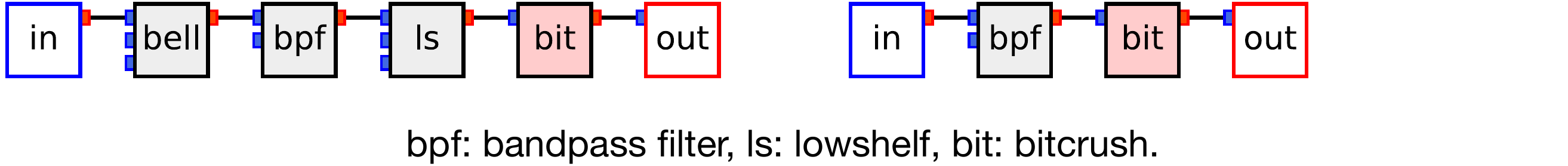}}\vspace{-1.7mm} \\
        \caption{Inference results from the proposed method \ding{194} on the singing voice effect estimation task (left: ground-truth, right: prediction).} 
  \label{fig:singing-est} 
  \vspace{-3.7mm}
\end{figure}

\noindent \textbf{Performance Analysis.} On the \emph{seen} dry speaker/kit sets, the \ding{194} proposed model  reports the node error rate of $0.215$ and $0.335$ for the singing and drum task, respectively, indicating that the perfect reconstruction of the prototype graph is rare.  
Yet, audio rendered from the estimated graph can be perceptually close to the reference, reporting the MUSHRA score of $76.6_{\pm 4.6}$ and $69.8_{\pm 5.4}$. 
On the \emph{unseen} dry speaker/kit sets, the evaluation results are degraded in most metrics, confirming that the graph estimation from unseen sources is challenging. We note that the majority of the errors come from either (i) the wrong order of processors (see Figure \ref{fig:drum-est} and \ref{fig:singing-est-1}) or (ii) some processors which destroy the signal and make preceding processors harder to notice (see Figure \ref{fig:singing-est-2}). 
Finally, to check the difficulty of the blind estimation, we trained the same model but with \ding{195} dry sources also provided as input by concatenating it with the reference across the channel axis. Indeed, the estimation performance improves by a noticeable margin, indicating that extracting the graph-relevant information solely from the reference is a challenging task.

\noindent \textbf{Blind estimation strategy comparison.} We compared our \ding{194} token-by-token approach with the conventional \ding{196} node-by-node decoding method \cite{li2018learning}. While this model uses the same TokenGT backbone, it estimates the next node type first and then performs an edge prediction task using the transformer outputs. It showed similar or slightly worse performance overall compared to our model \ding{194}, and it had the drawback of having a high invalid graph rate. Finally, we tried a \ding{197} single-stage generation method; we decoded the node/edge parameters along with the categorical types. Since the transformer only has access to the decoded intermediate graph, its parameter estimation performance was much worse than the two-stage approach, resulting in higher MSS-full loss and lower MUSHRA score.

\vspace{-2mm}
\section{Discussion}
\vspace{-2mm}

\noindent \textbf{Summary.}
We integrated the audio processing graph structure with deep learning methods. We first synthesized the graph-reference pair data, discussed its characteristics, and built a blind estimation system with the off-the-shelf neural network components. We found that the token-by-token generation is an effective method for our graph structure, and the two-stage approach separately treating the connectivity information and processor parameters is beneficial. 

\noindent \textbf{Future Works.}
(i) Currently, the synthetic data remain in the toy example level; more diverse processors and input source signals could be desirable. 
(ii) Many real-world audio processing graph structures allow multi-edges and cycles; relaxing our single-edge DAG constraint could allow more expressive processing capabilities. 
(iii) The evaluation results showed that encoding the references from the unseen source distribution is challenging; we need an improved method for encoding only the processing-relevant information in a disentangled manner.
(iv) The proposed graph decoder uses the default transformer with sinusoidal encodings; further performance improvement
might be obtained by explicitly injecting graph connectivity information.
(v) While we used ground-truth prototypes as input to train the parameter estimation task, at the inference time, the model uses decoded graphs, which are different from the originals in most cases. Furthermore, while not all parameters are equally important for perceptual similarity, we ignored such aspects.
Since most processors we used are known to be differentiable    \cite{neuralbiquads, steinmetz2022styletransfer, lee2022dar}, end-to-end training with audio-domain objectives could be possible and beneficial for alleviating such issues. 
(vi) With the differentiable processors, we can combine them with neural audio processors, allowing us to balance between interpretability/controllability and expressibility.
(vii) Finally, extending the current blind estimation framework to other applications,  e.g., automatic processing \cite{martinez2022automatic, koo2022remaster, stasis2017audio} and style transfer \cite{steinmetz2022styletransfer}, is a promising research direction.

\bibliographystyle{ieeebib-abbr}
\bibliography{refs}

\end{document}